# The Lorentz Transformation Sign Ambiguity and Its Relation to Measured Faster-than-c Photon Speeds[*]

by

Robert J. Buenker
Bergische Universität Wuppertal, Fachbereich C-Mathematik und Naturwissenschften,
Gaussstr. 20, D-42119 Wuppertal, Germany

**Abstract**

It is shown that time intervals $\Delta t'$ measured for photons moving with speed $v > c$ can be of the same sign for all observers according to special relativity, thereby avoiding any violation of Einstein causality. Previous assertions to the contrary have led to unnecessarily complicated interpretations of experiments which indicate that single photons do travel with faster-than-c speeds in regions of anomalous dispersion.

**I.     Introduction**

Shortly after the publication of Einstein's original paper on special relativity in 1905 [1], it was pointed out by Wien [2] that a possible contradiction to the new theory might occur for light passing through a region of anomalous dispersion.  Since the refractive index n can be less than unity under these conditions, it was argued that both the phase and group velocities of light can be greater than c, the speed of light in free space.  Continuing to the present day, it has been widely assumed that special relativity precludes the possibility of faster-than-c speeds, so in 1907 detailed arguments were presented by Sommerfeld [3] to show that the above experimental phenomenon can be reconciled with this theory as long as the corresponding signal and energy velocities do not exceed v = c.  In the last decade new experimental evidence has been presented [4,5] which shows quite clearly that single photons do propagate with v > c speeds in the neighborhood of absorption lines, but there has been a general reluctance to accept this result at face value [4,6], again because it is thought to be inconsistent with special relativity.

It is important to note that one of the objections to faster-than-c speeds is not applicable to the case of anomalous dispersion of light, and is in fact the only one actually mentioned by Einstein in his original work [1].  He pointed out that particles with non-zero proper mass would experience a singularity in their kinetic energy when accelerated to v = c and thus ruled out speeds at or above this value for any such entity, but by the same argument that this restriction does not apply to photons by virtue of their vanishing proper mass.  There is another objection, however, namely the assertion that Einstein causality must be violated according to special relativity whenever v > c because of a consequence of the Lorentz transformation [7].  This second difficulty is potentially relevant for photons and implies that the time order of two events can be opposite for two observers moving with different relative speeds, something which is justifiably considered to be strictly unphysical and thus to stand in contradiction to special relativity.

**II.    Phase Relationships and Einstein Causality**

In view of the lingering uncertainties in the interpretation of the recent experiments dealing with the anomalous dispersion phenomenon, it is well to reexamine the latter objection to faster-than-c speeds. The Lorentz transformation equation for space and time intervals $\Delta x$ and $\Delta t$ are given below for two observers moving at relative speed u in the x direction:

$$\Delta x'(u) = \varepsilon_x(u)\, \gamma(u)\, (\Delta x - u\, \Delta t) \tag{1}$$
$$\Delta t'(u) = \varepsilon_t(u)\, \gamma(u)\, (\Delta t - \frac{u}{c^2}\, \Delta x),$$

where $\gamma = (1 - \frac{u^2}{c^2})^{-1/2}$. The quantities $\varepsilon_x$ and $\varepsilon_t$ must be chosen to be consistent with the relativistic invariance condition for the two observers in their respective (primed and umprimed) inertial systems, namely

$$\overline{\Delta x'(u)}^2 - c^2\, \overline{\Delta t'(u)}^2 = \overline{\Delta x}^2 - c^2\, \overline{\Delta t}^2, \tag{2}$$

whereby the corresponding spatial intervals in the transverse directions for the two observers are taken to be equal in arriving at the above expression. This requirement leads to the conditions, $\varepsilon_x^2 = \varepsilon_t^2 = 1$, and thus to an ambiguity in the respective signs of these two quantities. This uncertainty is normally removed by noting that the above equations must reduce to the Galilean transformation for small relative speed u. Consequently, both $\varepsilon_x$ and $\varepsilon_t$ are assigned a value of +1, and by virtue of the physical requirement of continuous variation in both $\Delta x'(u)$ and $\Delta t'(u)$, it is argued that this result must hold quite generally for all accessible relative speeds $u < c$.

In the case of present interest the speed v of photons in a dispersive medium with a group index of refraction $n_g$ is defined as

$$v = \frac{\Delta x}{\Delta t} = \frac{c}{n_g}. \qquad (3)$$

Substitution in eq. (1) gives

$$\begin{aligned}
\Delta x' &= \varepsilon_x(u)\, \gamma(u)\, \Delta x\, (1 - \frac{u}{v}) \\
&= \varepsilon_x(u)\, \gamma(u)\, \Delta x\, (1 - \frac{n_g u}{c}) \\
\Delta t' &= \varepsilon_t(u)\, \gamma(u)\, \Delta t\, (1 - \frac{uv}{c^2}) \\
&= \varepsilon_t(u)\, \gamma(u)\, \Delta t\, (1 - \frac{u}{n_g c}),
\end{aligned} \qquad (4)$$

where again $\varepsilon_x$ and $\varepsilon_t$ are shown explicitly as in eq. (1). The conventional Einstein causality argument [7] then goes as follows. If $n_g < 1$, the quantity in parentheses in the expression for $\Delta t'$ can be negative for $c > u > n_g c$. By assuming as above that $\varepsilon_t = 1$, it is concluded that in this range of u, $\Delta t'$ must have the opposite sign as $\Delta t$, a clear violation of Einstein causality.

The latter argument overlooks an important point, however. Since $\Delta t'$ vanishes in eq. (4) for $u = n_g c$, it is no longer required on continuity grounds that $\varepsilon_t = +1$ for this value of u. All that is required physically is that $\Delta t'$ itself be a continuous function of u, and this objective can be accomplished with either choice of sign for $\varepsilon_t(u)$ at the critical speed $u = n_g c$.

In Fig. 1 both $\Delta x'$ and $c \Delta t'$ are given as a function of $\beta = \dfrac{u}{c}$ for a dispersive medium with $n_g = 0.5$. In this diagram $\varepsilon_x(u)$ always has a value of +1 but $\varepsilon_t(u)$ changes discontinuously from +1 to -1 at $\beta = 0.5$ (the light source moves in the opposite direction as the light itself from the vantage point of the observer). The relativistic invariance condition of eq. (2) is everywhere satisfied and a violation of Einstein causality does not occur, i.e. $\Delta t'$ is never opposite in sign to $\Delta t$. Most importantly, both $\Delta x'$ and $\Delta t'$ are continuous functions of u with the above choices for $\varepsilon_x(u)$ and $\varepsilon_t(u)$, so that all physical constraints are satisfied.

One other consequence of the discontinuity in $\varepsilon_t$ at $\beta = n_g$ is that the derivative $d\Delta t'/d\beta$ is not defined at this relative speed, but there is no compelling argument against this result. The observer simply measures a monotonically decreasing time interval as $\beta$ is gradually increased up to the critical value of $n_g$, at which point $\Delta t'$ vanishes (Fig. 1). As $\beta$ is further increased, the measured time interval begins to increase monotonically, always with the same sign as $\Delta t$. The spatial interval $\Delta x'$ also reaches a minimum value at $\beta = n_g$ and retains the same sign over the entire range of accessible relative speeds.

There remains one other conceivable objection to faster-than-c speeds, namely the fact that the measured velocity $v' = \dfrac{\Delta x'}{\Delta t'}$ has a singularity at $\beta = n_g$ by virtue of the vanishing of $\Delta t'$ at this relative speed (see Fig. 1). It is only a point singularity, however, which means that the observer would have to be moving at exactly the critical relative speed $\beta = n_g$ to measure other than finite values for $v'$ according to the present theory. It is important to recall that the arguments under discussion only apply to photons moving through an anomalously dispersive medium. This implies that the range over which unlimited speeds can be observed is quite restricted, since it is a practical impossibility to construct such a medium with a uniform group index of refraction which extends over more than a short distance.

The latter considerations are significant in another context, however. There has been speculation [8] that particles with imaginary proper masses (tachyons) might exist which would necessarily travel with faster-than-c speeds through free space. If such a particle were observed, special relativity again demands that there be a critical relative speed u ($\frac{c^2}{v}$) for which $\Delta t'$ vanishes and thus the measured speed of the tachyon would be infinite in the corresponding inertial system. Since the particle could attain this speed in free space, there is nothing in special relativity to prevent its being observed simultaneously in parts of the universe which are many light years apart. Although one can't fully rule out such a prospect, it is far less difficult to imagine that an observer moving at close to the speed of light relative to a laser source would measure a null value for the elapsed time taken by a single photon to cover a distance of only a few μm [4,5]. It also should be noted that the same argument that has been used above for space-time intervals can also be applied to energy-momentum four-vectors. Hence, by proper choice of the sign employed in the corresponding Lorentz transformation it is also possible to avoid the conclusion [8] that negative-energy states are the inevitable consequence of faster-than-c tachyon speeds.

With the above choice of signs for the Lorentz transformation it is seen that the velocity of the faster-than-c photon is in the *same* direction for all observers regardless of their relative speed u. This result is thus in sharp contrast to what is obtained when $\Delta t'$ is allowed to change sign as the critical speed $u = n_g c$ is surpassed, in which case $\frac{\Delta x'}{\Delta t'}$ changes abruptly from $+\infty$ to $-\infty$ at this point. Since the relative speed of the observer is always less than the photon's speed $c/n_g$ for $n_g < 1$ it must be expected that the perceived direction of the motion will always be the same, so the situation depicted in Fig. 1 is by far the more intuitively reasonable of the two possibilities. The fact that the photon's speed is observed to increase with $\beta$ up to the critical value of $n_g$, that is, as the observer moves more quickly away from the light source, is tied up with the fact that, although $\Delta x'$ and $\Delta t'$ are both decreasing with $\beta$ in this range (Fig. 1), the latter is decreasing faster on a fractional basis, ultimately reaching a null value at $\beta = n_g$. For slower-than-c speeds it is $\Delta x'$ that is decreasing faster and hence the measured speed of the object decreases with $\beta$ over the entire range of accessible relative speeds. In this case the key velocity is $u = v = \Delta x/\Delta t$ [see eq. (1)], for which $\Delta x'$ vanishes and the object appears to reverse direction as the relative speed of the observer is further increased.

### III. Relation to Experiment

Once it is realized that by appropriate choice of signs in the Lorentz transformation it is possible to avoid a violation of Einstein causality in the theoretical description of faster-than-c speeds, it is no longer necessary to look for other than a straightforward interpretation of the experimental results obtained in regions of anomalous dispersion [4,5]. In other words, the individual photons observed in these experiments really do travel with $v > c$. At least there is no evidence to the contrary, either theoretical or experimental. Each photon is a separate entity with a definite speed. Many of them are absorbed in such experiments [4,6] and thus do not reach the detector, but there is every reason to believe that those that are transmitted are moving at the speed actually measured for them. This conclusion is also consistent with recent measurements of photon velocity distributions in dispersive media *vis-a-vis* those in air [9]. Time-correlated single photon counting (TCSPC) detection indicates that the $\delta$-function character of the velocity distribution of photons emanating from a laser source is left unchanged as the light passes from air through water, and hence that the speed of *each* photon is determined solely by the value of the group refractive index $n_g$ in any given medium. Carrying this result over to the case of media with $n_g < 1$ simply leads one to expect what has been observed experimentally, namely that each transmitted photon is actually moving with $v > c$.

**IV. Energy and Signal velocity**

As a final remark, it is important to consider the arguments given by Sommerfeld and Brillouin [2,10,11] regarding the definitions of energy and signal velocity and their relation to the more conventional terms, phase and group velocity. Their conclusion was that the way to reconcile special relativity with the fact that the group velocity of light exceeds c in regions of anomalous dispersion ($n_g < 1$) is to realize that the wave front and the signal associated with the wave motion propagate at different speeds which can never exceed c. This position needs to be reassessed in light of the present finding that purely mathematical considerations eliminate the apparent inconsistency between special relativity and such experimental results. Since Einstein's work on the photoelectric effect [12], it has been recognized that the energy of light waves is carried by individual photons. On an elemental level, therefore, the energy velocity must be identical with that of the photon itself. Moreover, a single photon can be a signal, opening doors or sending an alarm, for example, so again in this sense, the photon velocity and the signal velocity cannot be distinguished from one another. It might be argued that a photon with an energy close to that of an absorption line is not a reliable signal because of the high probability of its being absorbed prior to reaching a suitable detector, but simply increasing the intensity of the laser beam makes it possible to effectively eliminate any time delay that would result because the earliest photons were not transmitted.

**V. Conclusion**

In summary, the main advantage of making the choice of signs advocated in the present work for the Lorentz transformation for particles moving with v > c is that it allows one to interpret all relevant experimental results in a simple and straightforward manner without violating Einstein causality and without making any change in the postulatory structure of the theory of special relativity.

**Acknowledgment**

This work was supported in part by the Deutsche Forschungsgemeinschaft within the Schwerpunkt Programm *Theorie relativistischer Effekte in der Chemie und Physik schwerer Elemente.* The financial support of the Fonds der Chemischen Industrie is also hereby gratefully acknowledged.

**Figure Captions**

**Fig. 1.**

Schematic diagram showing the variation of the spatial and time intervals $\Delta x'$ and $c\,\Delta t'$ as a function of the relative speed $\beta = \dfrac{u}{c}$ of the observer for a photon traveling at twice the speed of light in free space ($n_g = 0.5$) in the inertial system with $\beta = 0$. Their ratio (measured velocity) is also shown. Note that $\Delta t'$ is always positive in this diagram by virtue of the choice of sign $\varepsilon_t(u)$ in the corresponding Lorentz transformation of eq. (4).

**Fig.1**

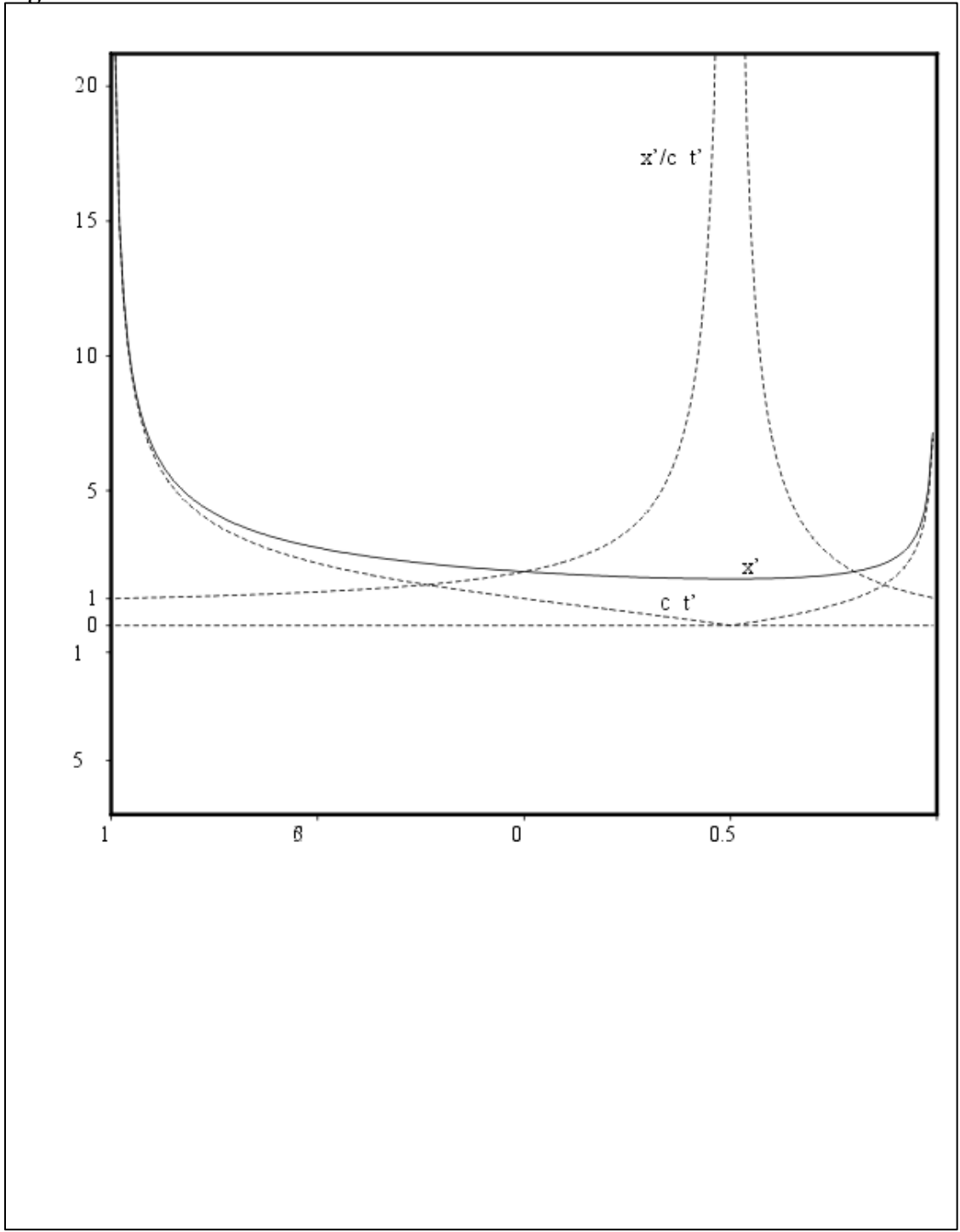